\documentclass[12pt,preprint]{aastex}

\shorttitle{A Universal Power-law Profile in Elliptical Galaxies}
\shortauthors{Chae}

\begin{document}

\title{A Universal Power-law Profile of Pseudo-Phase-Space Density-like 
Quantities in Elliptical Galaxies}

\author{Kyu-Hyun Chae}
\affil{Department of Astronomy and Space Science, Sejong University, 
 98 Gunja-dong, Gwangjin-Gu, Seoul 143-747, Republic of Korea}
\email{chae@sejong.ac.kr}

\begin{abstract}
We study profiles of mass density, velocity dispersion (VD), and their 
combination using  $\sim 2000$ nearly spherical and rotation-free SDSS 
galaxies. For observational stellar mass density $\rho_{\star}(r)$ 
 we consider a range of dark matter (DM) 
distribution $\rho_{\rm{DM}}(r)$ and VD anisotropy $\beta(r)$ to investigate
radial stellar VD $\sigma_{\rm\star r}(r)$ using the spherical Jeans equation. 
 While mass and VD profiles vary appreciably depending on DM distribution 
 and anisotropy, the pseudo-phase-space density-like combination  
$\rho(r)/\sigma_{\rm\star r}^3(r)$ with total density
$\rho(r)= \rho_{\star}(r)+\rho_{\rm{DM}}(r)$ is nearly universal.
In the optical region the minus of its logarithmic slope 
 has a mean value of $\langle\chi\rangle\approx 1.86$--$1.90$ 
with a galaxy-to-galaxy rms scatter of $\approx 0.04$--$0.06$, 
which is a few times smaller than that of $\rho(r)$ profiles. 
The scatter of $\chi$ can be increased by 
invoking wildly varying anisotropies that are, however, less likely because 
they would produce too large a scatter of line-of-sight VD profiles.  
As an independent check of this 
universality we analyze stellar orbit-based dynamical models of 15 ETGs of 
Coma cluster provided by J. Thomas. Coma ETGs, with $\sigma_{\star\rm{r}}(r)$ 
replaced by the rms velocity of stars $v_{\star\rm{rms}}(r)$ 
including net rotation, exhibit a similar 
universality with a slope of $\chi= 1.93\pm 0.06$.
Remarkably, the inferred values of $\chi$ for ETGs match well the slope 
$\approx 1.9$ predicted by N-body simulations of DM halos. 
We argue that the inferred universal nature of $\rho(r)/\sigma_{\rm\star r}^3(r)$ 
 cannot be fully explained by equilibrium alone, 
implying that some astrophysical factors conspire and/or it reflects a 
fundamental principle in collisionless formation processes.

\end{abstract}


\section{Introduction}

 A useful approach in studies of galaxies is 
to look for any regularity or universality in the structure and then study
astrophysical factors and/or principles responsible for it.
Regarding galaxy profiles, until now much attention has been paid to 
possibility of a universal density profile of galaxies 
(e.g.\ \citealt{Per,Ger,LP,Koo,Rem}). However, 
there is yet no reliable theoretical or numerical prediction on a 
universal mass profile of galaxies as existing models are flawed or incomplete 
and observational results on galactic mass profile are not converging well.
 On the observational side, there have been suggestions for approximately 
universal isothermal density profile $\rho(r)\propto r^{-2}$ for both 
spiral/disk galaxies (e.g.\ \citealt{Rub,Per}) and early-type 
 galaxies (ETGs; e.g.\ \citealt{Ger,HB,Koo}). 
However, some recent results give a mean
 profile steeper-than-isothermal in ETGs \citep{CBK,Son,Bol}. 
Moreover, the density slope $\gamma\equiv -d\ln\rho(r)/d\ln r$ may vary with 
radius in both spirals (e.g.\ \citealt{Per}) and ellipticals 
(e.g.\ \citealt{CBK}) or with redshift \citep{Son,Bol}, and
the intrinsic galaxy-to-galaxy scatter is significant even within $R_{\rm e}$
 $0.1< \sigma_\gamma \la 0.2$ (e.g.\ \citealt{Koo,CBK}).
 On the theoretical side,
the classical isothermal model by \cite{Lyn}
 has several problems including infinite mass, indefinitely growing
entropy \citep{LW} and non-transitive nature \citep{AL} and the applicability 
and usefulness of equilibrium statistical mechanics to gravitationally-bound
collisionless systems are still being investigated 
(e.g.\ \citealt{Tre,WN,HK,BW,PG}). Hydrodynamic simulations of galaxy formation
in the halo are used to study how the pristine 
(i.e.\ primitive before galaxy formation) halo is 
modified during the course of galaxy formation (e.g.\ \citealt{Duf,Gne,Macc}). 
However, physics involved in hydrodynamic simulations is so complex 
that a robust prediction on the mass profile is challenging.  
Therefore, the case for a universal density profile of galaxies as a dynamical 
attractor or fixed point in dynamical evolution (see, e.g., \citealt{LP}) 
appears weak at present.

As first noticed by \cite{TN}, N-body simulations of hierarchical cold dark
matter (CDM) halo formation show that the combination 
$\rho_{\rm pDM}(r)/\sigma_{\rm pDM}^3(r)$ of  mass density
$\rho_{\rm pDM}(r)$ and velocity dispersion (VD) $\sigma_{\rm pDM}(r)$ of the
pristine DM halo, which is called pseudo phase-space (PPS) density because it 
has the dimension of the phase-space density (or distribution 
function) but is not a true measure of it (e.g., \citealt{AB,SS,Vas}), 
is closer to universal than $\rho_{\rm pDM}(r)$
and can be well described by a scale-free power-law profile 
with slope $\chi_{\rm pDM}\equiv-d\ln[\rho_{\rm pDM}(r)/\sigma_{\rm pDM}^3(r)]/
d\ln r\approx 1.9$ over three orders of magnitude in radius 
(e.g.\ \citealt{Asc,Aus,WW,Vas,Nav10,Lud}).
It appears that $\rho_{\rm pDM}(r)/\sigma_{\rm pDM}^3(r)$ rather than 
$\rho_{\rm pDM}(r)$ offers a more powerful route to the universal nature of 
pristine halos. 
In this context, significant efforts have
been made to investigate the physical origin of the universality, scale-free 
nature and slope value of the $\rho_{\rm pDM}(r)/\sigma_{\rm pDM}^3(r)$-profile 
(e.g.\ \citealt{TN,Aus,Bar,WW,Vas,Lud,LC}) and their implication for 
the structure of pristine DM halos (e.g.\ \citealt{Aus,DM}). 

 In this work we examine mass and VD profiles of galaxies and their 
combination using SDSS elliptical galaxies. It is argued that 
there exists a universal profile of PPS density-like quantities in elliptical 
galaxies, akin to the universal PPS density profile of pristine DM halos, 
with an intrinsic scatter smaller than the observed scatter of 
mass density profiles.
 
\section{Samples of Early-type Galaxies and Their Models}

 We consider a sample of nearly spherical galaxies that allows a
relatively straightforward analysis. The galaxy sample is drawn 
from the Sloan Digital Sky Survey (SDSS: \cite{CBK} and references theirin)  
and contains $\sim 2000$ nearly spherical 
(surface brightness ellipticity $< 0.15$) and disk-less (disk mass is within
the measurement error of the bulge mass) galaxies at redshifts $z \la 0.25$. 
Each galaxy is assumed to be in a dynamical 
equilibrium state satisfying the spherical Jeans equation \citep{BT} given by 
 \begin{equation}
\frac{d[\rho_{\star}(r) \sigma_{\rm \star r}^2(r)]}{dr} 
+ 2 \frac{\beta(r)}{r} [\rho_{\star}(r) \sigma_{\rm \star r}^2(r)]
= - G \frac{\rho_{\star}(r) M(r)}{r^2},
\label{eq:Jeans}
\end{equation}
where $\rho_{\star}(r)$ is the volume stellar mass distribution, 
$M(r)$ is the total mass within $r$,
 $\sigma_{\rm \star r}(r)$ is the radial stellar VD, and 
$\beta(r)$ is the VD anisotropy given by
$\beta(r)=1 -\left[\sigma_{\star \theta}^2(r) + \sigma_{\star \phi}^2(r)\right]/
 \left[ 2 \sigma_{\rm \star r}^2(r) \right]$ 
where $\sigma_{\star \theta}(r)$ and $\sigma_{\star \phi}(r)$ are tangential VDs
 of stars in spherical coordinates. 
 The Jeans equation can be used to obtain a VD profile 
$\sigma_{\rm \star r}(r)$ for an observationally inferred $\rho_{\star}(r)$ 
if $\rho_{\rm DM}(r)$ and $\beta(r)$ are known or specified. 
Conversely, if a line-of-sight (LOS) VD profile 
$\sigma_{\rm \star los}(r)$ is measured for some radial range, 
the Jeans equation can be used to derive $\rho_{\rm DM}(r)$ and $\beta(r)$ 
along with $\sigma_{\rm \star r}(r)$ for the observed radial range.
Without $\sigma_{\rm \star los}(r)$ for SDSS galaxies, 
we take the former approach and consider a range of 
$\rho_{\rm DM}(r)$ and $\beta(r)$. 
 For DM distribution we consider both the case of no DM and
 assigning a generalized Navarro-Frenk-White (gNFW) halo 
$\rho_{\rm gNFW}(r)\propto r^{-\alpha} (1+r/r_{\rm s})^{-3+\alpha}$ using a wealth of 
empirical information, as in \cite{CBK}, including statistical distribution of 
LOSVD profiles within $\sim R_{\rm e}$ from well-studied ETGs 
 and statistical properties and relations of ETGs and DM halos.
 We consider both 
constant anisotropy and radially varying anisotropy using a function of
the form $\beta(r)=\beta_1/(1+r_1^2/r^2)+\beta_2/(1+r_2^2/r^2)$, which is a 
combination of two Osipkov-Merritt-type \citep{BT} models 
but allows an extremum at a finite radius. 
We take anisotropy values randomly from an observed distribution
described in section~2.7 of \cite{CBK}. For the case of radially varying 
anisotropy the mean anisotropy for $r< R_{\rm e}$ and the anisotropy at infinity
take a common value drawn from the observed distribution and 
$0<r_1 < 0.5 R_{\rm e}$ and $r_1 <r_2 <  R_{\rm e}$ are assigned so that an 
extremum occurs within $R_{\rm e}$. Radially varying anisotropies with an 
extremum (or extrema) within $R_{\rm e}$ result often from dynamical modeling of
ETGs (e.g.\ \citealt{Ger,Tho07}).
 
We also consider a small number of ETGs of the Coma cluster for which 
$\sigma_{\rm los}(r)$ has been individually measured up to $> R_{\rm e}$ and thus 
$\rho_{\rm DM}(r)$ and $\beta(r)$ along with $\sigma_{\rm \star r}(r)$ have been 
derived from the data \citep{Tho07,Tho09}. Specifically,
two-component axisymmetric galaxy models are used and 
a best-fit model is calculated 
through a maximum entropy implementation of 
Schwarzschild's stellar orbit superposition technique
  fitting the observed LOSVD profile.
The detailed modeling results are provided by J. Thomas (private communication).
The Coma sample consists of 15 ETGs selected from 19 ETGs \citep{Tho07,Tho09}. 
Our criterion for this selection is that the mean of the radial ranges along 
the major and minor axes over which LOSVD profiles were observed extends beyond 
$R_{\rm e}/2$. Seven of them (GMP 144, 282, 1750, 3510, 3792, 5279 and 5975) 
are ellipticals without significant disks. The rest are two 
ellipticals (GMP 2440, 3958) possessing significant disks and six lenticulars
(GMP 756, 1176, 1990, 2417, 3414, and 4928). The excluded galaxies are
GMP 2921, 3329, 4822 and 5568, three of which are cD/D galaxies. 
Because streaming motion can be significant for lenticulars and 
ellipticals with non-negligible disks, we consider an average 1-dimensional 
second velocity moment given by
\begin{equation}
 v_{\rm \star rms}(r)=\left[ \frac{\sigma_{\rm \star r}^2(r)+\sigma_{\star \theta}^2(r)
  +\sigma_{\star \phi}^2(r)+\bar{v}_{\star\phi}^2(r)}{3}\right]^{1/2} 
= \left[ \sigma_{\rm \star}^2(r)+\frac{1}{3}\bar{v}_{\star\phi}^2(r) \right]^{1/2},
\label{eq:vrms}
\end{equation} 
where $\bar{v}_{\star\phi}(r)$ is the net intrinsic rotation speed of stars at 
$r$ taking into account the inclination of each Coma galaxy
(\citealt{Tho07}; J. Thomas, private communication).
 Notice that $v_{\rm \star rms}(r)=\sigma_\star(r)$ if $\bar{v}_{\star\phi}(r)=0$.

\section{Profiles of  Density $\rho(r)$, Velocity Dispersion 
 $\sigma_{\rm \star r}(r)$ and  $\rho(r)/\sigma_{\rm \star r}^3(r)$}

If a quantity $\rho(r)/\sigma_{\rm \star r}^\epsilon(r)$ should follow a universal
 power-law profile at least for some radial range, then we expect a good 
correlation between slopes $\gamma\equiv -d\ln \rho(r)/d\ln r$ and  
$\eta\equiv -d\ln \sigma_{\rm \star r}(r)/d\ln r$ for some radial range. Here the 
exponent $\epsilon$ is unknown but a theoretically interesting value is 
$\epsilon=3$ for which the quantity has the dimension of phase space density.
If we write $\rho(r)/\sigma_{\rm \star r}^\epsilon(r)\propto r^{-a}$, then we expect
$\gamma=a + \epsilon \eta$ for some radial range. To test this hypothesis
 we calculate values of $\gamma$ and $\eta$ and check if there is a correlation.
For the SDSS sample we consider a fixed radial range of
$0.1R_{\rm e}< r < R_{\rm e}$ as this range is most relevant for the modeling
results \citep{CBK}. For the Coma cluster sample we use the mean of the 
radial ranges of observed LOSVD profiles along the major and minor axes,
which varies from galaxy to galaxy but is typically 
$0.1R_{\rm e}\la r \la 1.5R_{\rm e}$. 
 For Coma galaxies having net rotations $\sigma_{\rm\star r}(r)$ is replaced 
by $v_{\rm\star rms}(r)$ (Equation~\ref{eq:vrms}).
 
\begin{figure}
\includegraphics[angle=-90,scale=.65]{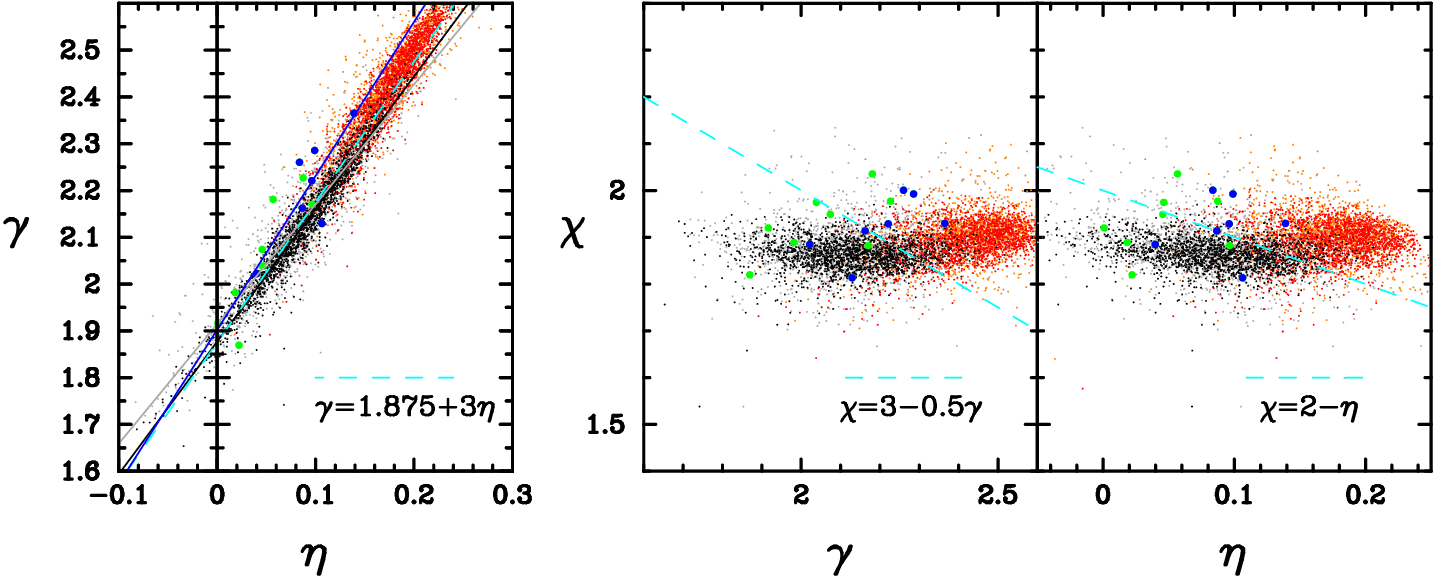}
\caption{Distribution of the slopes $\gamma$ for $\rho(r)$, 
$\eta$ for $\sigma_{\star \rm{r}}(r)$ , and $\chi$ for 
$\rho(r)/\sigma_{\star \rm{r}}^3(r)$ for $0.1 R_{\rm e}< r < R_{\rm e}$
of $\sim 2000$ nearly spherical and rotation-free SDSS galaxies.
 Black and gray (red and orange) points are respectively for constant and 
radially varying anisotropies with (without) DM halos. 
 Blue and green points are the values 
for $0.1 R_{\rm e} \la r \la 1.5 R_{\rm e}$, respectively of
elliptical and lenticular (S0) galaxies of Coma cluster, with 
 $\sigma_{\star \rm{r}}(r)$ replaced by $v_{\star \rm{rms}}(r)$.
Solid lines are the least-square fit relations. The cyan dashed line
$\gamma = 1.875+3\eta$ is the prediction by the classical self-similar 
spherical infall model \citep{Bert,FG}. The cyan dashed lines 
$\chi=3-0.5\gamma$ and $\chi=2-\eta$ are the approximate predictions by
Jeans (or virial) equilibrium equation.
 \label{slopeplane}}
\end{figure}

The left panel of Figure~\ref{slopeplane} displays the correlation between
 $\gamma$ and $\eta$. Good correlations are found independently from 
both samples. The least-square fit relations are 
$\gamma \approx 1.87 + 2.83\eta$ (constant anisotropy),  
 $\gamma \approx 1.91 + 2.63\eta$ (varying anisotropy) for the SDSS sample 
and $\gamma \approx 1.90 + 3.29\eta$ for the Coma sample. 
Similar results are found even for the case of no DM
[i.e.\ $\rho(r)=\rho_\star(r)$] in which stars are self-gravitating.
These results imply that a PPS density-like
combination with $\epsilon=3$ is expected to be (close to) universal with a 
power-law exponent of $\approx -1.9$, which interestingly matches well the 
slope $-1.875$ predicted by the classical self-similar spherical-infall model 
\citep{Bert,FG}.
The middle and right panels of Figure~\ref{slopeplane} display the distributions
 of $\chi\equiv -d\ln [\rho(r)/\sigma_{\rm \star r}^3(r)]/d\ln r$ against 
$\gamma$ and $\eta$. The slope $\chi$ is not correlated with either $\gamma$ or 
$\eta$. This lack of correlation of $\chi$ is what would be expected if 
$\rho(r)/\sigma_{\rm \star r}^3(r)$ should be universal. 

\begin{figure}
\includegraphics[angle=-90,scale=.60]{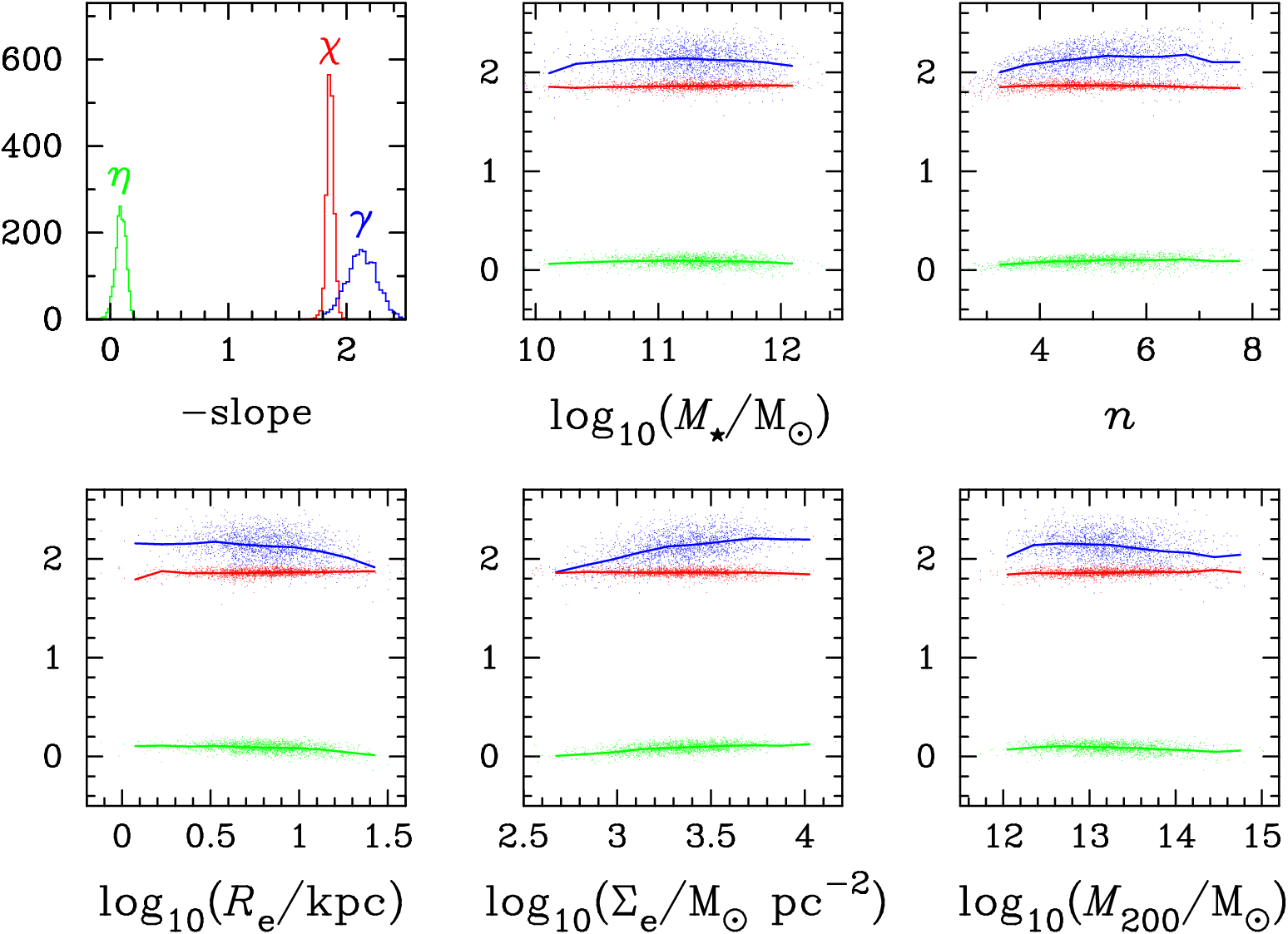}
\caption{Distribution of $\gamma$, 
$\eta$, and $\chi$ for the galaxies shown in 
Figure~\ref{slopeplane} against various parameters. $\chi$ has the
smallest scatter and is not correlated with any parameters while 
$\gamma$ and $\eta$ show some (anti-)correlations with
$R_{\rm e}$ and $\Sigma_{\rm e}$.
 \label{slope_e}}
\end{figure}

The large number of galaxies in the SDSS sample allows us to investigate 
how slopes are distributed as a function of various parameters.
Figure~\ref{slope_e} displays the distributions of $\gamma$, $\eta$
and $\chi$ against stellar mass $M_\star$, S\'{e}rsic \citep{Ser} index $n$, 
effective radius $R_{\rm e}$, projected stellar mass density 
$\Sigma_{\rm e}\equiv (M_\star/2)/(2\pi R_{\rm e}^2)$ within $R_{\rm e}$ 
and the host halo virial mass $M_{200}$. Only the results with constant 
anisotropies are displayed here but the case of varying anisotropy is
qualitatively similar.
 It is evident that $\chi$ is little 
correlated with any galactic or halo parameter. In contrast $\gamma$ varies 
systematically in particular with $R_{\rm e}$ and $\Sigma_{\rm e}$. 
Slope $\eta$ also exhibits some systematic trends but to a lesser degree. 
The mean and rms scatter of the slopes are:
$\langle\gamma\rangle=2.13$, $s_{\gamma}=0.13$; 
$\langle\eta\rangle=0.09$, $s_{\eta}=0.05$; and, 
 $\langle\chi\rangle=1.86$, $s_{\chi}=0.04$. 
Remarkably, slope $\chi$ has much smaller scatter compared with $\gamma$ as can
 be clearly seen in the histogram (top left panel) of Figure~\ref{slope_e}.
 The scatter of $\chi$ increases up to $0.06$--$0.08$ for general ETGs of 
any ellipticity for constant anisotropies 
and further increases by $0.01$--$0.02$ for the adopted radially varying 
anisotropy. The scatter of $\chi$ can be increased even 
up to or beyond the scatter of $\gamma$ by considering wildly
varying anisotropies which however predict too large scatters of LOSVD profiles.
The small scatter of $\chi$ is also suggested by the Coma cluster sample for
which we have: $\langle\gamma\rangle=2.13$, $s_{\gamma}=0.14$; 
$\langle\eta\rangle=0.07$, $s_{\eta}=0.04$; and, 
 $\langle\chi\rangle=1.93$, $s_{\chi}=0.06$. 
Therefore, it appears that $\rho(r)/\sigma_{\rm \star r}^3(r)$ 
(or $\rho(r)/v_{\rm \star rms}^3(r)$) is
closer to universal than $\rho(r)$ at least within $\sim R_{\rm e}$.

\begin{figure}
\includegraphics[angle=-90,scale=.60]{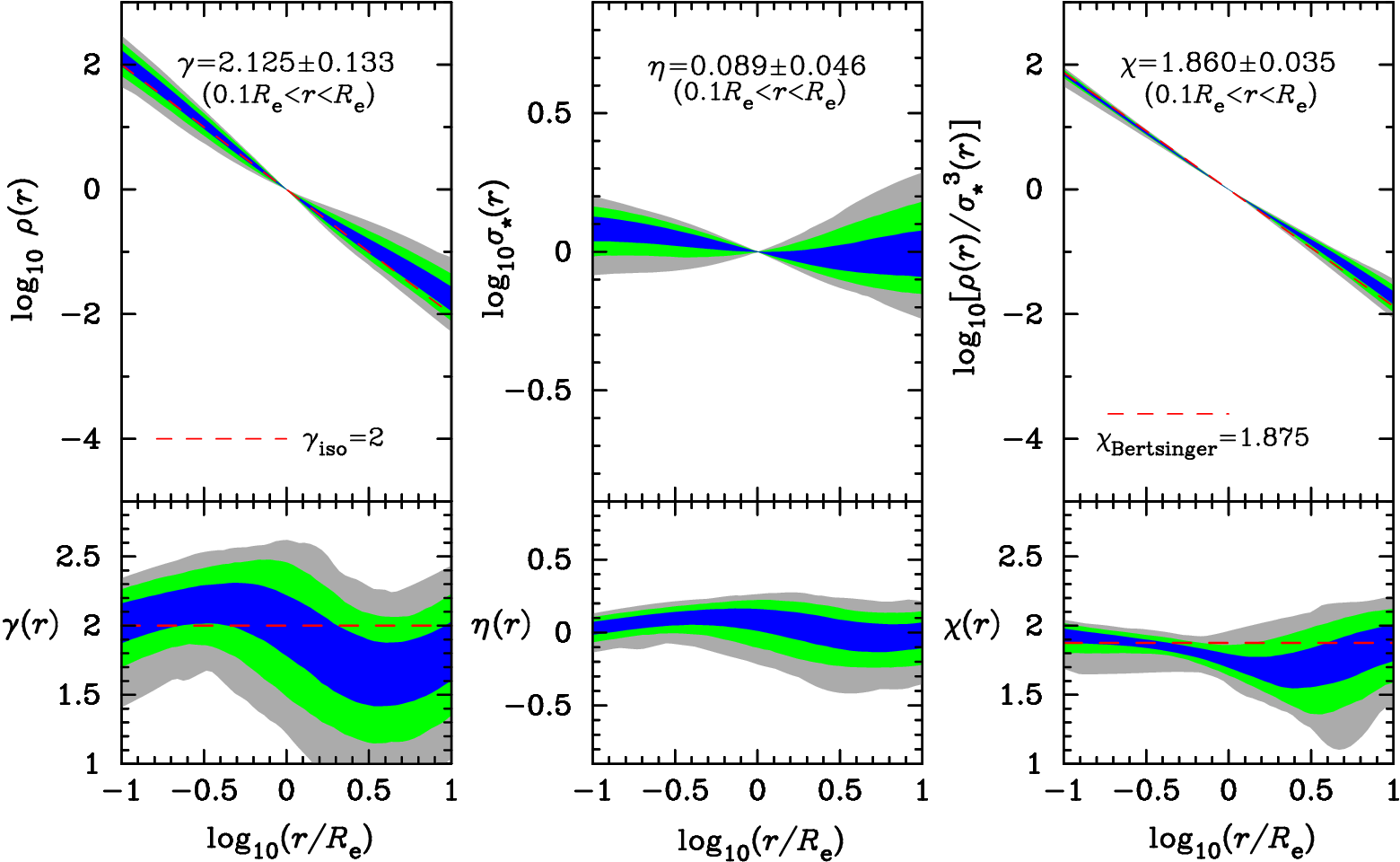}
\caption{Profiles of $\rho(r)$, $\sigma_{\star{\rm r}}(r)$ and 
$\rho(r)/\sigma_{\star{\rm r}}^3(r)$  for the galaxies 
shown in Figure~\ref{slopeplane}.  
All quantities are normalized to the values at $r=R_{\rm e}$. 
Blue, green and gray regions contain 68\%, 95\% and 99.7\% of
galaxies respectively. Bottom panels show profiles of negative slopes. 
 \label{profs}}
\end{figure}

Figure~\ref{profs} displays the predicted
profiles up to $10 R_{\rm e}$ based on the SDSS sample. 
Compared with $\rho(r)$ $\rho(r)/\sigma_{\rm \star r}^3(r)$  is
much closer to a universal power-law profile. Figure~\ref{profComa} displays 
the profiles of 15 Coma cluster ETGs directly constrained by the individually
 measured LOSVDs. These profiles independently indicate that for ellipticals
$\rho(r)/\sigma_{\rm \star r}^3(r)$ is closer to universal than $\rho(r)$.
What is even more striking from the Coma profiles is that
all ETGs including lenticulars appear to
 follow the universal profile when $v_{\rm \star rms}(r)$ 
(Equation~\ref{eq:vrms}) rather than  $\sigma_{\rm\star r}(r)$ is used.
 
Unlike $\rho(r)/\sigma_{\rm \star r}^3(r)$, the quantity
$\rho_\star(r)/\sigma_{\rm \star r}^3(r)$ does not exhibit a universality for a
realistic galaxy embedded in a DM halo. The scatter of the slope 
$-d\ln [\rho_\star(r)/\sigma_{\rm \star r}^3(r)]/d\ln r$ is 
even larger than that of $\gamma$. This property is 
rooted in the fact that stellar motions (DM motions as well) are not 
self-gravitating but governed by the total mass distribution, 
as can be seen in Equation~(\ref{eq:Jeans}). 
In a pristine DM halo in which DM motions are governed by
its own gravity the PPS density $\rho_{\rm pDM}(r)/\sigma^3_{\rm pDM}(r)$ 
exhibits a universality (e.g.\ \citealt{TN,Aus,Lud,Nav10}).
However, in `real' DM halos hosting galaxies the PPS density of DM 
particles $\rho_{\rm DM}(r)/\sigma_{\rm DM}^3(r)$ is not universal or power-law 
as can be shown using our models. This property was also noticed in 
cosmological hydrodynamic simulations (e.g.\ \citealt{Zem}). 

\section{A Simple Analysis of Equilibrium}

Let us analyze the spherical Jeans equation to see if the universality 
of $\rho(r)/\sigma_{\rm \star r}^3(r)$ can be a natural outcome of equilibrium.
Using Equation~(\ref{eq:Jeans}) 
one can express a PPS density-like quantity with the radial VD as follows:
\begin{equation}
\frac{y(x)}{z^3(x)} = \frac{1}{\kappa}
\frac{\gamma_\star(x)+2\eta(x)-2\beta(x)}{x^2 z(x)} \left(1-2\eta(x)+
\frac{d\ln[\gamma_\star(x)+2\eta(x)-2\beta(x)]}{d\ln x}\right),
\label{eq:Jeansn}
\end{equation}
where $x\equiv r/r_0$, $y(x)\equiv\rho(r)/\rho(r_0)$, 
$z(x)\equiv\sigma_{\rm \star r}(r)/\sigma_{\rm \star r}(r_0)$, and 
$\kappa\equiv 4\pi G\rho(r_0) r_0^2/\sigma_{\rm \star r}^2(r_0)$ with an 
arbitrary reference radius $r_0$ (see \cite{TN}). 
In Equation~(\ref{eq:Jeansn}) the following symbols are used:
$\gamma_\star(x)\equiv-d\ln y_\star(x)/d\ln x$ with 
$y_\star(x)\equiv \rho_\star(r)/\rho_\star(r_0)$, and 
$\eta(x)\equiv-d\ln z(x)/d\ln x$.

The last factor in the large parenthesis of the right-hand side of 
Equation~(\ref{eq:Jeansn}) is $\approx 1$. In the numerator of the second
factor of the right-hand side 
$\gamma_\star$ ($ \approx 0.9+1.9 (r / R_{\rm e})^{1/n} \ga 2$ 
for $r > 0.1R_{\rm e}$ assuming the S\'{e}rsic profile) is dominating as 
$|\eta| \la 0.2$ and $|\beta| \la 0.5$ in most cases. 
For some radial range, e.g.\ $0.1 R_{\rm e}< r < R_{\rm e}$, we may write
$z(x) = x^{-\eta}$ to a good approximation. We may also write the universal
PPS density-like profile as $y(x)/z^3(x)=x^{-\chi}$. Then we are left with
 $\chi \approx 2-\eta$. Similarly, if we write $y(x) = x^{-\gamma}$ for 
$0.1 R_{\rm e}< r < R_{\rm e}$, we obtain $\chi\approx 3-0.5\gamma$. 
These relations can be used to predict a mean value 
$\langle\chi\rangle \approx 1.9$ by empirical values 
of $\langle\eta\rangle\approx 0.1$ or $\langle\gamma\rangle\approx 2.1$.
They also imply that the intrinsic scatter of $\chi$ is comparable to
that of $\eta$ but one half of that of $\gamma$. These predictions of the mean
and the intrinsic scatter are consistent with the results from our galaxy 
models. However, as shown in the middle and right panels of 
Figure~\ref{slopeplane} 
these approximate relations are not obeyed by our galaxy models. 
Our galaxy models show no correlation of $\chi$ with either $\eta$ or $\gamma$.
Therefore, although equilibrium can predict the mean and the scatter of
the slope of $\rho(r)/\sigma_{\rm \star r}^3(r)$, equilibrium itself cannot 
generically explain its universal nature seen in our models. For simple galaxy 
models equilibrium demands correlations of $\chi$ with $\gamma$ and $\eta$.
The seen universality free of such correlations implies that some other 
astrophysical factors and/or principles are at work in (our models of) 
real galaxies. 

For a galaxy with net rotation one may assume that stars in a local volume 
centered at $r$ satisfy a virial equation $v_{\rm \star rms}^2(r) = - \Phi(r)$ 
(similar to the equation for the whole galaxy)
 where $v_{\rm \star rms}(r)$ is given by Equation~(\ref{eq:vrms}) and 
$\Phi(r)$ is the gravitational potential. 
From this equation similar relations of
$\chi$ with $\eta$ and $\gamma$ can be obtained.

\section{Discussion}

 A range of DM distribution and VD anisotropy in elliptical
galaxies have been considered for investigating galaxy mass and VD profiles.
 The PPS density-like combination $\rho(r)/\sigma_{\rm \star r}^3(r)$ 
(or $\rho(r)/v_{\rm \star rms}^3(r)$ for general ETGs)  
appears to be closer to universal than $\rho(r)$. Moreover, the
universality of $\rho(r)/\sigma_{\rm \star r}^3(r)$ of ellipticals closely 
parallels the universality of the PPS density of pristine DM halos 
(e.g.\ \citealt{Aus,Nav10,Lud}) with the 
values of slope $\chi$ agreeing well.
The PPS density is known to be scale-free over three
orders of magnitude in radius for N-body simulated halos.
Our galaxy models indicate a power-law profile of 
$\rho(r)/\sigma_{\rm \star r}^3(r)$ in the optical region, and 
extrapolation from our models suggests a scale-free profile over two orders 
of magnitude in radius.  It appears then natural to suggest that the PPS 
density-like quantity $\rho(r)/\sigma_{x{\rm r}}^3(r)$
for any collisionless component $x$ in dynamical equilibrium follows a 
universal power-law profile at least for some radial range.

\begin{figure}
\includegraphics[angle=-90,scale=.62]{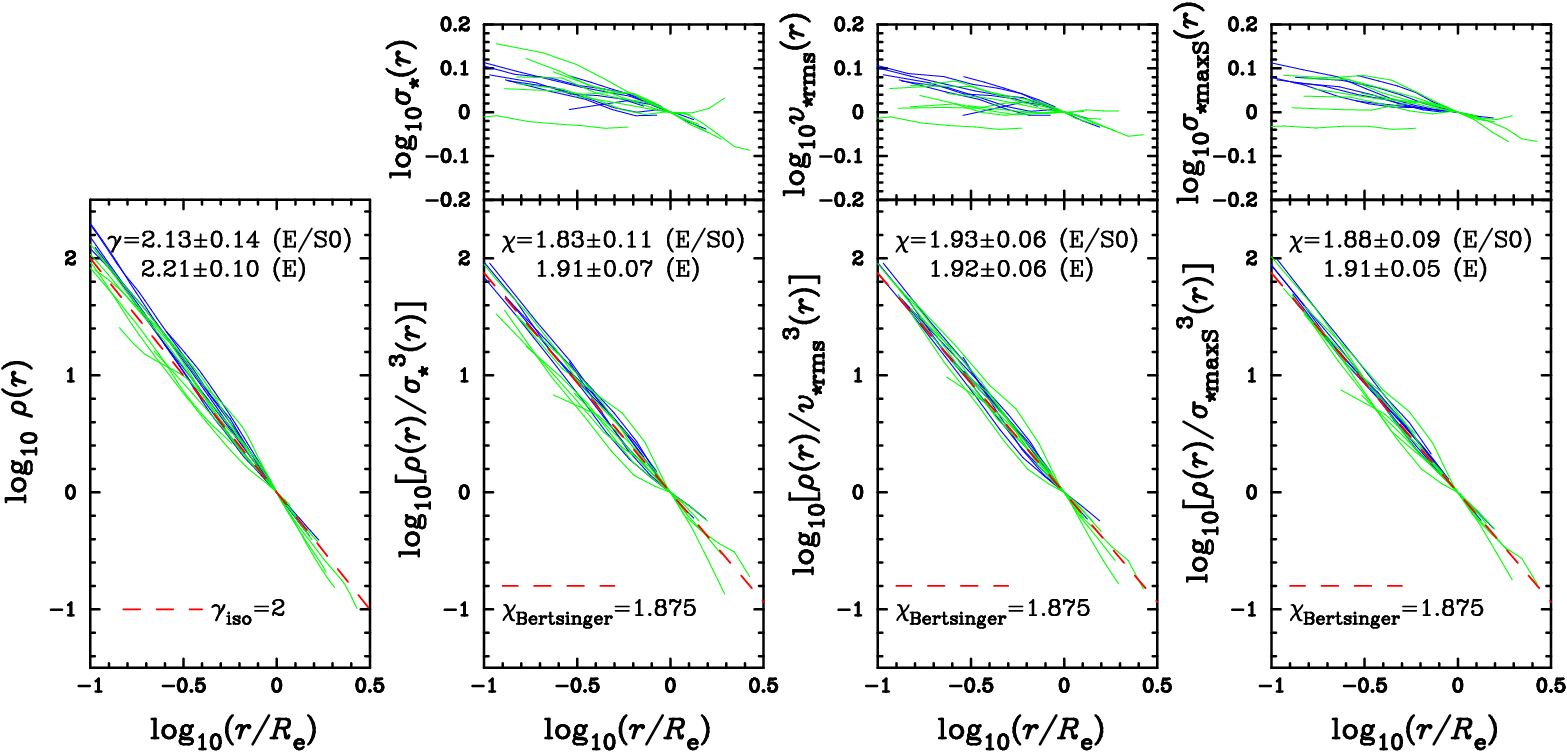}
\caption{Profiles as in Figure~\ref{profs} but for 15 ETGs of Coma cluster 
shown in Figure~\ref{slopeplane}. Note that $v_{\star{\rm rms}}(r)$ in general
 differs from $\sigma_{\star}(r)$ as some ETGs (in particular lenticulars) have 
non-negligible net rotations. $\sigma_{\star{\rm maxS}}(r)$ is the velocity 
dispersion in models obtained by maximizing entropy but not fitting well 
observed LOSVDs.
 \label{profComa}}
\end{figure}

Examination of Jeans (or virial) equation shows that equilibrium can
predict $\chi=3 - 0.5 \gamma$ giving 
a mean slope of $\langle\chi\rangle\approx 1.9$ for an empirical 
$\langle\gamma\rangle\approx 2.1$ in the optical region. However, equilibrium 
cannot generically predict a power-law profile or the lack of correlation 
of $\chi$ with galactic properties. 
What astrophysical factors and/or physical principles can be attributed to
the shared universality of $\rho(r)/\sigma_{x{\rm r}}^3(r)$ in N-body simulated 
pristine halos and our ellipticals? Total mass profiles have no similarity 
between them. Moreover, the physical formation process is quite different. 
The pristine halo has formed through fast collapse, major and minor mergers, 
and accretion. Only dissipationless processes are involved. On the other hand, 
an ETG is believed to form through merging of disk galaxies and/or pre-existing 
ETGs. In the formation of the stellar disk dissipational physics plays the 
essential role and consequently the total density deviates from the NFW 
\citep{NFW} profile of the pristine halo and gets much steeper 
in the central region. The shared universality of 
$\rho(r)/\sigma_{x{\rm r}}^3(r)$ despite large differences
in mass profiles, structures and formation histories between pristine DM halos 
and ETGs suggests that the universality is not an outcome of specific 
astrophysical factors but reflects a fundamental physical 
principle/nature governing collisionless systems.
For Coma cluster ETGs maximum entropy models ignoring observed LOSVD
profiles, as shown in the right-most panel of Figure~\ref{profComa}, 
 resemble best-fit models in a statistical sense. 
This result is in line with the finding by \cite{Tho09a} that 
flattening of an ETG by stellar anisotropy maximizes the entropy for a given
density distribution. However, the relevance of maximum entropy principle for
collisionless systems under gravity is not well understood at present.
Perhaps, dynamical mixing/relaxation may be responsible for the universality 
(see, e.g., \citealt{Val}). 

While $\rho(r)/\sigma_{x{\rm r}}^3(r)$ appears to be universal, total mass density
$\rho(r)$ (not to mention the component density $\rho_x(r)$) is
determined by other physics. For example, purely radial collapse scenario by
\cite{Bert} gives $\rho(r)\propto r^{-2.25}$ while inclusion of non-radial 
motions can give a profile similar to the NFW (e.g.\ \citealt{Asc,Mac,Vog}).
In ETGs baryonic physics is responsible for the steep profile
$\rho(r)\propto r^{-\gamma}$ with $\gamma\sim 2.1$ in the optical region. 

\acknowledgments

The author is indebted to Jens Thomas for providing him with unpublished 
results on Coma cluster ETGs and useful discussions. 
The author thanks Andrey Kravtsov for useful discussions and helpful comments
on the manuscript. The author also would like to thank the anonymous referees 
 for useful comments. 

\bibliographystyle{mn2e}

\clearpage

\clearpage

\clearpage

\clearpage

\end{document}